# Micromagnetic simulations of sweep-rate dependent coercivity in perpendicular recording media

M. L. Plumer[1], M. D. Leblanc[1], J. P. Whitehead[1], and J. van Ek[2]
[1]*Department of Physics and Physical Oceanography, Memorial University of Newfoundland, St. John's, NL A1B 3X7, Canada*
[2]*Western Digital Corporation, San Jose, CA 94588, USA*

The results of micromagnetic simulations are presented which examine the impact of thermal fluctuations on sweep rate dependent coercivities of both single-layer and exchange-coupled-composite (ECC) perpendicular magnetic recording media. M-H loops are calculated at four temperatures and sweep rates spanning five decades with fields applied normal to the plane and at 45 degrees. The impact of interactions between grains is evaluated. The results indicate a significantly weaker sweep-rate dependence for ECC media suggesting more robustness to long-term thermal effects. Fitting the modeled results to Sharrock-like scaling proposed by Feng and Visscher [J. Appl. Phys. **95**, 7043 (2004)] is successful only in the case of single-layer media with the field normal to the plane.

## I. Introduction

Micromagnetic simulations have evolved into an essential tool for the evaluation of new designs and concepts in magnetic recording media. The most efficient and common means to assess the magnetic switching properties of storage media is through the examination of M-H hysteresis loops.[1] Modeling such experimental data through the use of Landau-Lifshitz-Gilbert (LLG) simulations is often used to estimate the microscopic parameters associated with granular media, such as the saturation moment ($M_s$), uniaxial magnetic anisotropy (K), inter-grain exchange (A) and various distributions in these parameters. However, it has long been recognized that the ns time scales involved in typical LLG simulations are very many orders of magnitude smaller than the typical measurement times of many minutes.[2,3] This is an important consideration at finite (e.g., room) temperatures where magnetic-grain switching is described by stochastic thermally assisted energy-barrier hopping involving an attempt frequency $f_0$. Despite its popularity, fitting microscopic LLG media parameters based on comparisons of T=0 K simulations with room-temperature experimental M-H loops cannot be considered reliable.

There has been considerable effort to understand the impact of M-H loop sweep rate (R) at finite temperatures on the coercivity ($H_c$). Many of these involve semi-analytic re-formulations of the Sharrock law, and all are restricted to the behavior of non-interacting particles, usually with the switching field applied only in the easy-axis direction.[3-5] Several other approaches include early work that combines an LLG calculation of the energy barriers with Monte Carlo (MC) simulations of the switching probability which allows for a completely numerical prediction of the long-time coercive force.[6] More recently, MC or LLG simulations have been combined with nudged elastic band model to estimate these effects on single particles or grains of interacting exchange spring bi-layers, with fields applied in various directions.[7,8] An alternative scheme has also been proposed which is based on Arrhenius-Néel scaling of short-time high-temperature, LLG simulations of M-H loops to estimate long-time room-temperature switching behavior.[9] Lacking in these studies is a comprehensive exploration of direct micromagentic modeling results for the sweep-rate and temperature dependence of $H_c$ for perpendicular recording media of current interest.

In the present work, LLG simulations of M-H loops performed on single-layer and exchange-coupled composite (ECC) granular perpendicular recording media from short time (~ 10 ns) to medium time (~ 1 ms) scales at four values of temperature. The field is applied either normal to the plane or at 45 degrees in an effort to illuminate field-direction dependent effects relevant to recording head fields. Simulations are also obtained with and without inter-grain interactions. These numerical results are fitted to the Arrhenius-Néel/Sharrock type scaling model proposed by Feng and Visscher.[5]

## II. Micromagnetic simulations

Simulations were performed using a commercial LLG simulator[10] with (for the single-layer case) 8x8x10 $nm^3$ grains on a 16x16 square lattice with periodic boundary conditions. Finite temperature effects were included through the usual Langevin stochastic term with an Euler integration routine, damping factor $\alpha$=0.1, and a time step of 0.2 ps. The saturation magnetization, perpendicular uniaxial anisotropy and weak inter-particle exchange were assigned mean values $M_{s1}$=500 emu/cc, $K_1=3.5\times10^6$ erg/cc, and $A_1$=0.05 μerg/cm, respectively. Note that at 300 K, these parameters give $K_1V/k_BT=58$. The magnetization and anisotropy were assumed to have a Gaussian distribution characterized by a standard deviation of 10%. In addition, the direction of the grain anisotropy axes were also given a distribution of $3^0$.

Magnetostatic interactions were included. These values are appropriate to model Co-Cr-Pt based recording media. We note that distributions in magnetic properties also serve to account for distributions in geometrical features, such as grain size (not explicitly included in the present model).

For ECC media the above parameters were used for the hard layer. A second thinner soft layer using 8x8x5 $nm^3$ grains was added to the simulation with $M_{s2}$=600 emu/cc, weaker (perpendicular) anisotropy $K_2=1.5\times10^6$ erg/cc and inter-grain exchange again set at $A_2$=0.05 μerg/cm (although it can be larger for some real media). Gaussian distributions as in the hard layer were also assumed. Exchange coupling between the layers was set at $J_{12}=2$ erg/$cm^2$, which corresponds to a moderate value of $A_{12}=J_{12}d_1/2$ = 0.8 μerg/cm, where $d_1$ is the thickness of the hard layer.[11]

In order to test the Arrhenius-Néel based scaling model of Feng and Visscher[5], M-H loops were calculated at four temperatures T=300, 600, 900 and 1200 K and at five sweep rates between R = $10^5 - 1^0$ Oe/ns. The field was applied either normal to the plane or at an angle of $45^0$ with a strength of $2\times10^4$ Oe in steps of 500 Oe. Note that in order to be consistent with the usual scaling assumptions, magnetic degrees of freedom internal to the grains were assumed to be frozen so that $M_s$ and K were taken to be temperature independent.[12,13]

In the case of single-layer media, four kinds of simulations were performed: (1) with all interactions included, (2) no magnetostatics, (3) no inter-particle exchange, and (4) no magnetostatics or inter-particle exchange.[7] For ECC media, only simulations with and without magnetostatic interactions were made. A comparison of these results allows for the evaluation of the impact of inter-particle interactions on the scaling of $H_c$ with T and R. Such interactions are omitted in analytic and semi-analytic Sharrock-like formulations of $H_c$(T,R).

## III. SINGLE-LAYER MEDIA

Fig. 1 illustrates some example (half) M-H loops for single-layer media with the field applied normal to the plane at 300 K with and without magnetostatic interactions. The general shape of the loops is consistent with experimental data on high anisotropy CoCrPt-based perpendicular media.[14] We note that usual LLG simulations solutions that are based on a tolerance criterion (typically about $10^{-4}$ in the magnetization vectors) yield loops that correspond to these results with R roughly between 1000 and 100 Oe/ns with a large $H_c$ ~ 9.5 kOe.. Omitting magnetostatic interactions from the simulations results in steeper slopes and larger coercivities, as has been previously observed in other magnetic systems.[7,12] Simulations run on PC workstations for the long-time sweep rates (R = 1 Oe/ns) took about four days to complete.

Corresponding results for the field applied at $45^0$ to the media plane are shown in Fig. 2. These loops exhibit the expected reduction in $H_c$ by about a factor of 3 and, notably, a reduced dependence on sweep rate. Omission of magnetostatic interactions has qualitatively the same general effect as in Fig. 1. These loops also show asymmetry on the approach to saturation between positive and negative field values, also a feature of the Stoner-Wohlfarth model.[15]

A summary of the sweep-rate dependence of $H_c$ with the field normal to the plane at 300 K for all four cases of inter-particle interactions omitted or not is shown in Fig. 3. Even at this moderate temperature, significant thermo-dynamic effects are seen with a 27% reduction in $H_c$ (all interactions included) over the five decades of R. Nearly logarithmic dependence is observed in all cases at the longer time scales (smaller R) with slopes which are roughly comparable in magnitude. For the curves corresponding to no magnetostatic interactions, there appears to be non-logarithmic dependence at the shorter time scales. If magnetostatic interactions are included, there appears to be little impact of removing inter-particle exchange (black vs green curves). Inter-particle exchange has a larger effect in the absence of magnetostatic interactions (red vs purple curves). The R-dependence of $H_c$ appears to be dominated by magento-crystalline anisotropy, as is the value of $H_c$ itself.

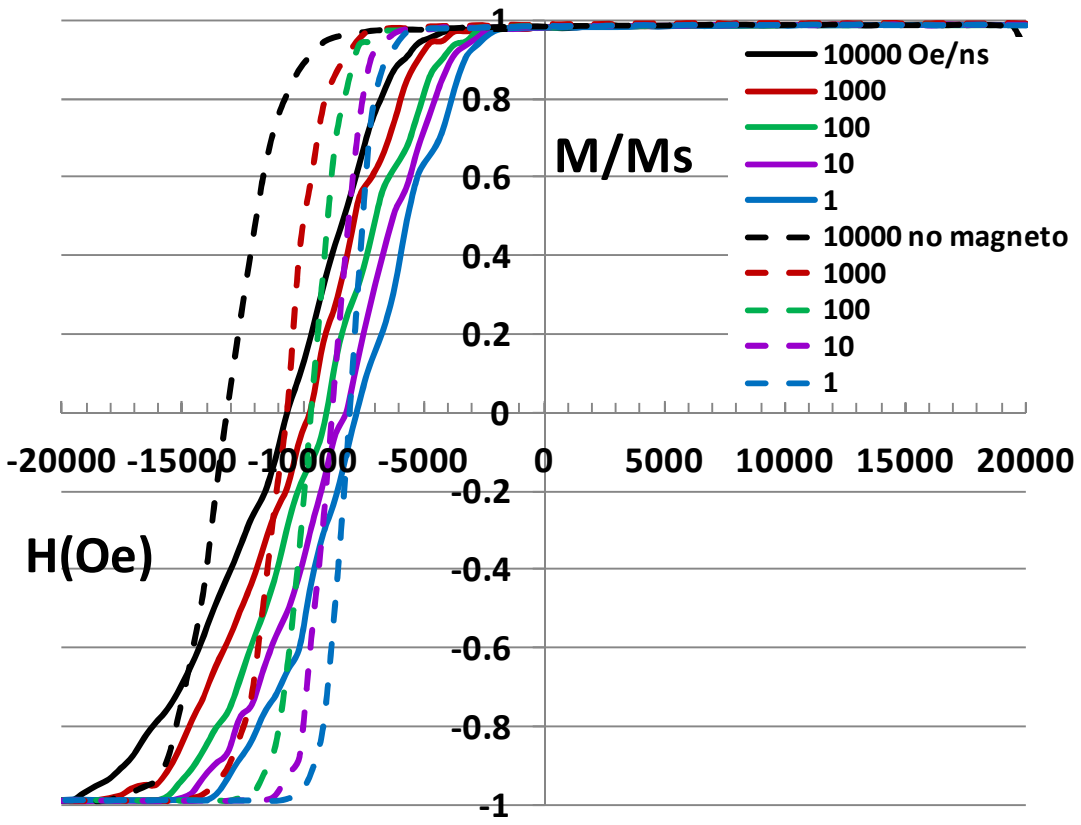

Fig. 1. Single-layer loops with the field normal to the media plane at 300 K with and without magnetostatic interactions included. Five different sweep rates were used, as indicated.

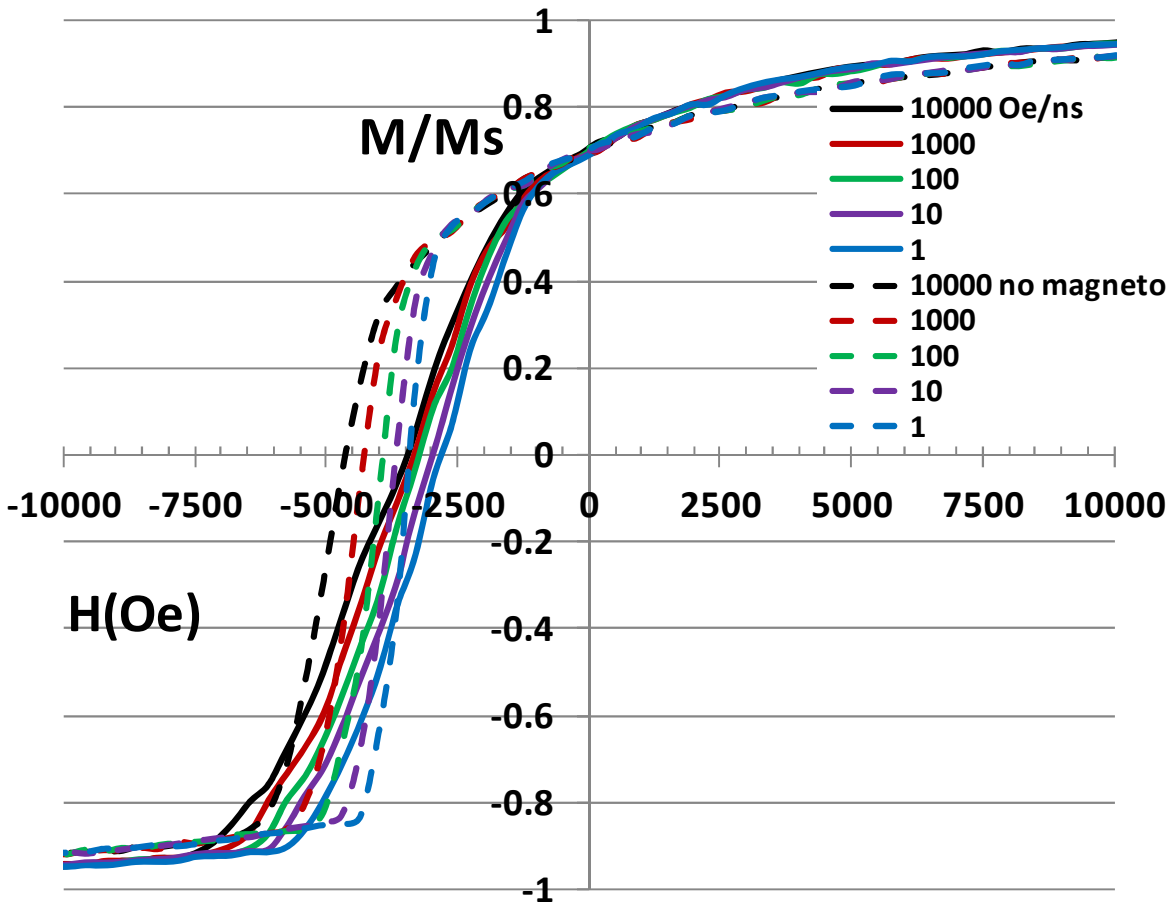

Fig. 2. As in Fig. 1 but with the field applied $45^0$ to the media plane.

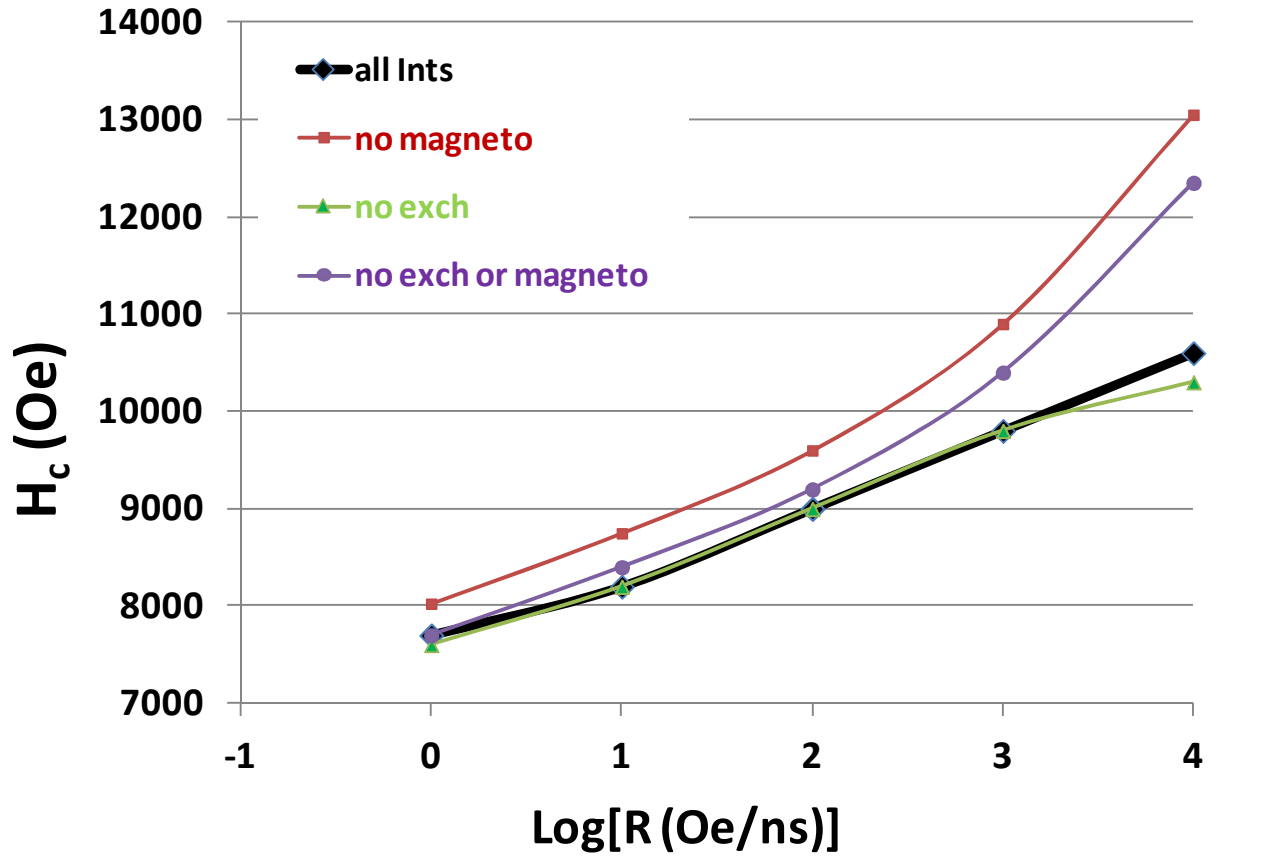

Fig. 3. Summary of coercivity vs logarithm of the sweep rate at 300 K with the field normal to the plane in single-layer media with various inter-particle interactions omitted as indicated.

Corresponding results for the field applied $45^0$ to the plane normal are shown in Fig. 4. In this case there is a very significant reduction in the impact of sweep rate down to 19% over the five decades of R values. As in Fig. 3, there appears to be little impact of inter-particle interactions on the rate of reduction. Note, however, that the overall effect of exchange interactions is negligible even in the absence of magnetostatic interactions. We believe this reflects the more dominant role played by domain rotation (compared to domain wall motion) as the field is applied further away from the easy axis.[15] Note also that in contrast with Fig. 3, the no-magnetostatic case (red curve) shows nearly logarithmic behavior even at the shorter time scales.

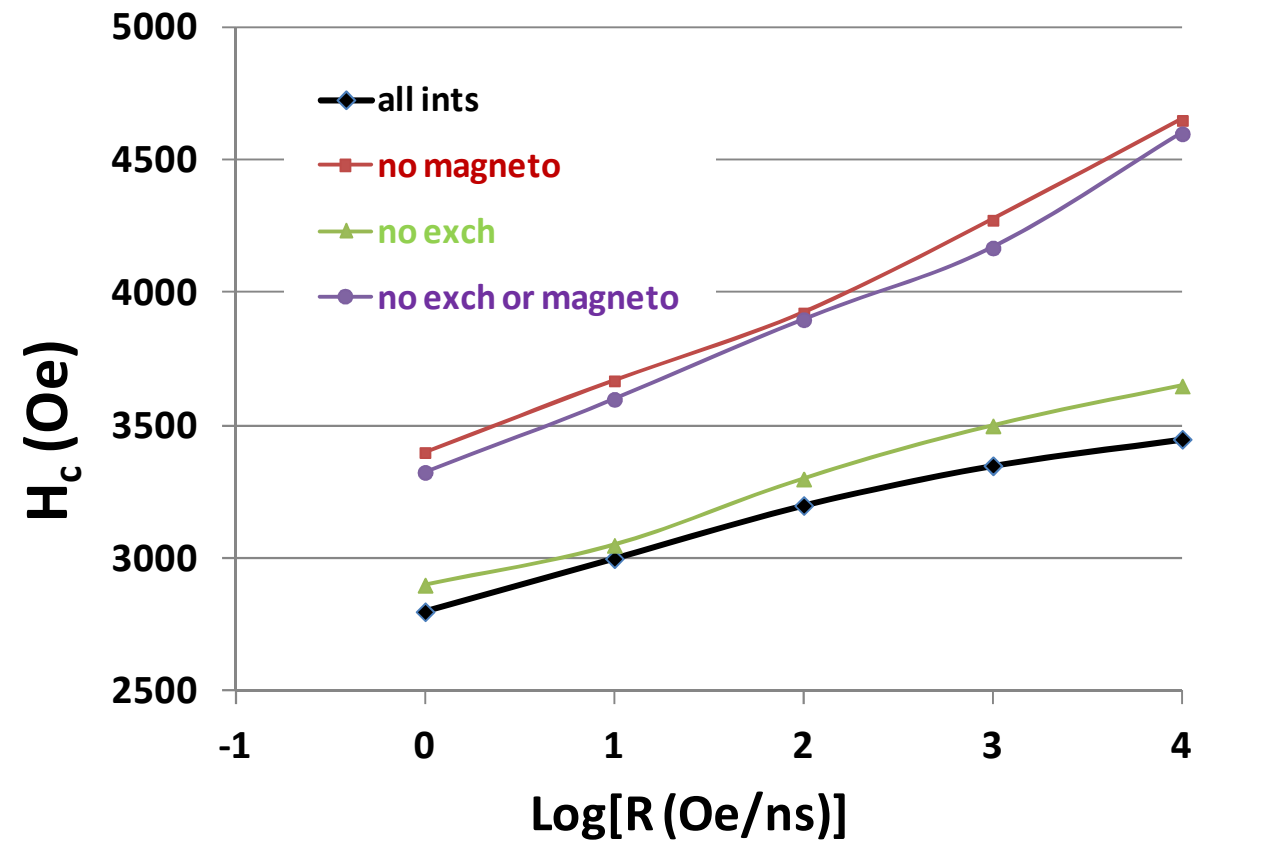

Fig. 4. As in Fig. 3 but with the field applied at $45^0$ to the media plane.

Fig. 5 summarizes the R dependence of $H_c$ at all four temperatures for the field normal to the media plane, with and without magnetostatic interactions (exchange is included in both cases). A significant increase in sensitivity to dynamic effects is seen at elevated temperatures, as expected from Arrhenius-Néel-like behavior. Note also that inter-particle interactions do not make a significant impact in the long-time behavior but there is not enough data to make a definitive conclusion on this point. Similar observations can be made regarding the simulation results summarized in Fig. 6 in the case of the field applied at $45^0$.

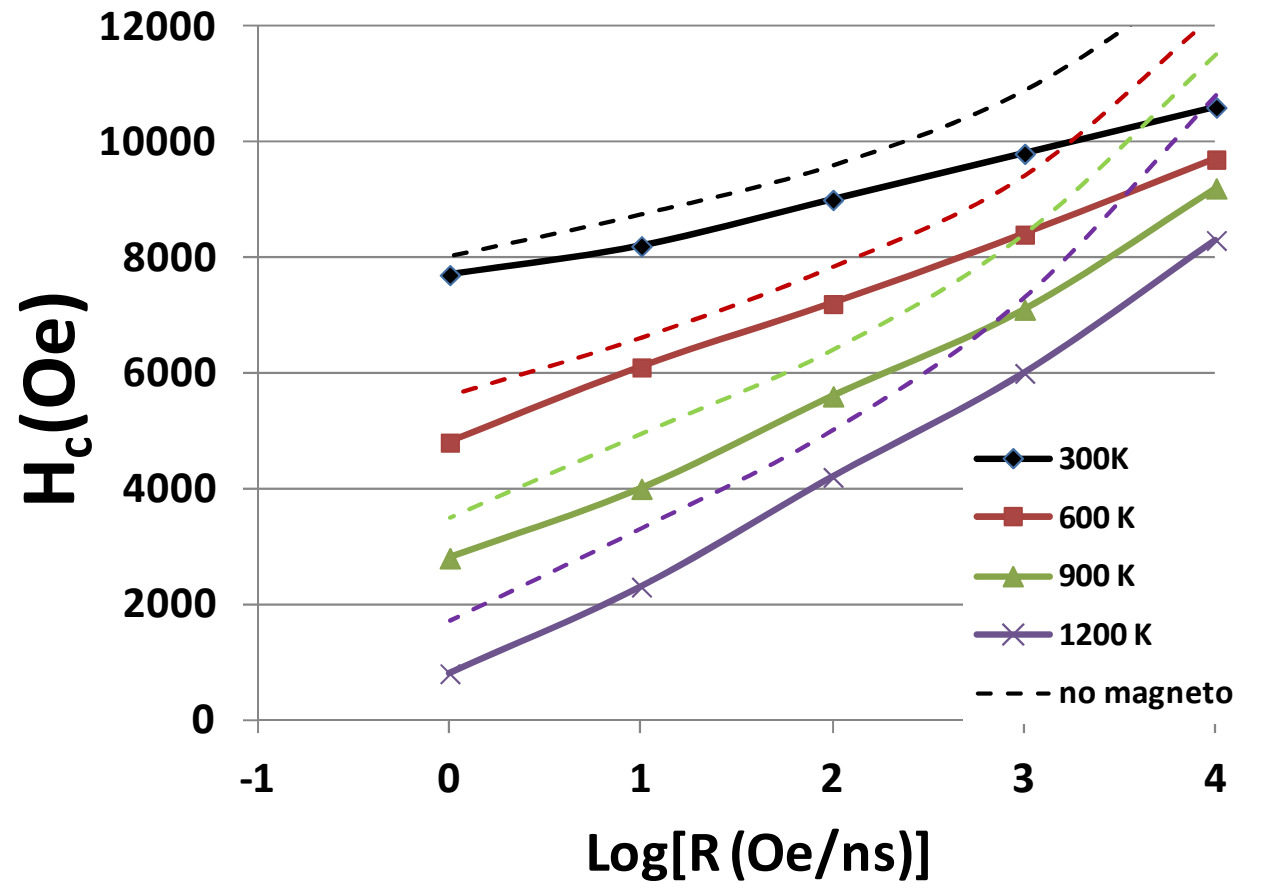

Fig. 5. Temperature dependence of $H_c$ vs Log[R] for the field applied normal to the plane in single-layer media.

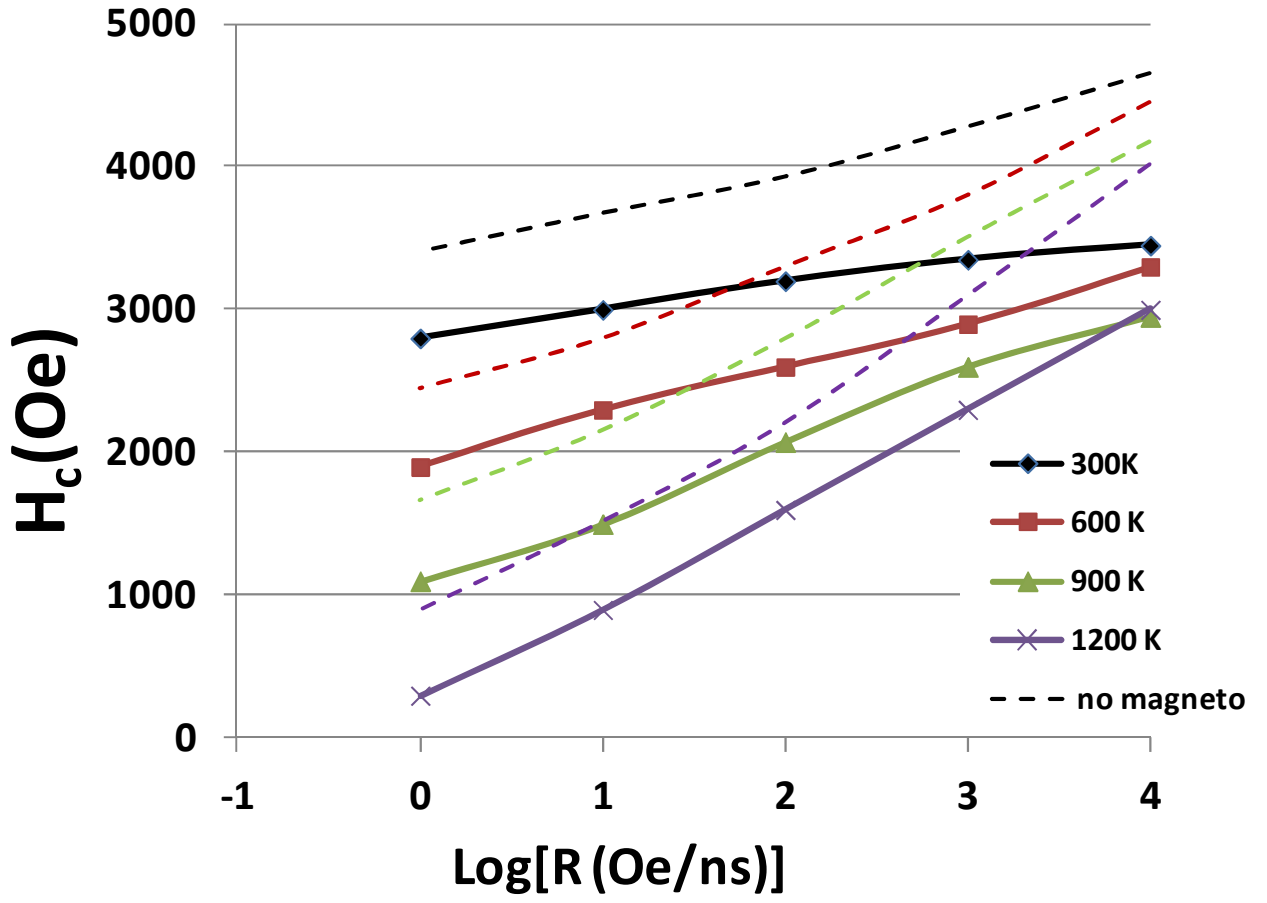

Fig. 6. As in Fig. 5 but with the applied field at $45^0$ to the media plane.

## IV. ECC MEDIA.

Commonly used in hard drives today, ECC media offers the advantage of similar thermal stability and recorded transition quality at a reduced switching field when compared to single-layer media.[14] With the larger number of magnetic parameters, micromagentic modeling of experimental loops is more challenging. Results corresponding to those of the previous section on single-layer case are presented here on ECC media. Loops at the longest sweep rate took about seven days to complete.

Fig. 7 shows loops at 300 K for the field applied along the easy axis direction at the five sweep rates, with and without the magnetostatic iteraction included. The overall coercivities are reduced , giving a nominal value of $H_c$ ~ 6 kOe compared with the single-layer case where $H_c$ ~ 9 kOe (for R between 1000 and 100 Oe/ns). The shapes of the loops are very similar to those of Fig. 1 and similar results have been reported for experimental loops, suggesting similar recording characteristics.[14] Loops with the field at $45^0$ shown in Fig. 8 exhibit only a factor of about two reduction in $H_c$ values compared to the case with the field normal to the plane (cf. a factor of three in the single-layer case). The impact of omitting magnetostatic interaction is similar to that seen in Fig. 2 in the single-layer case.

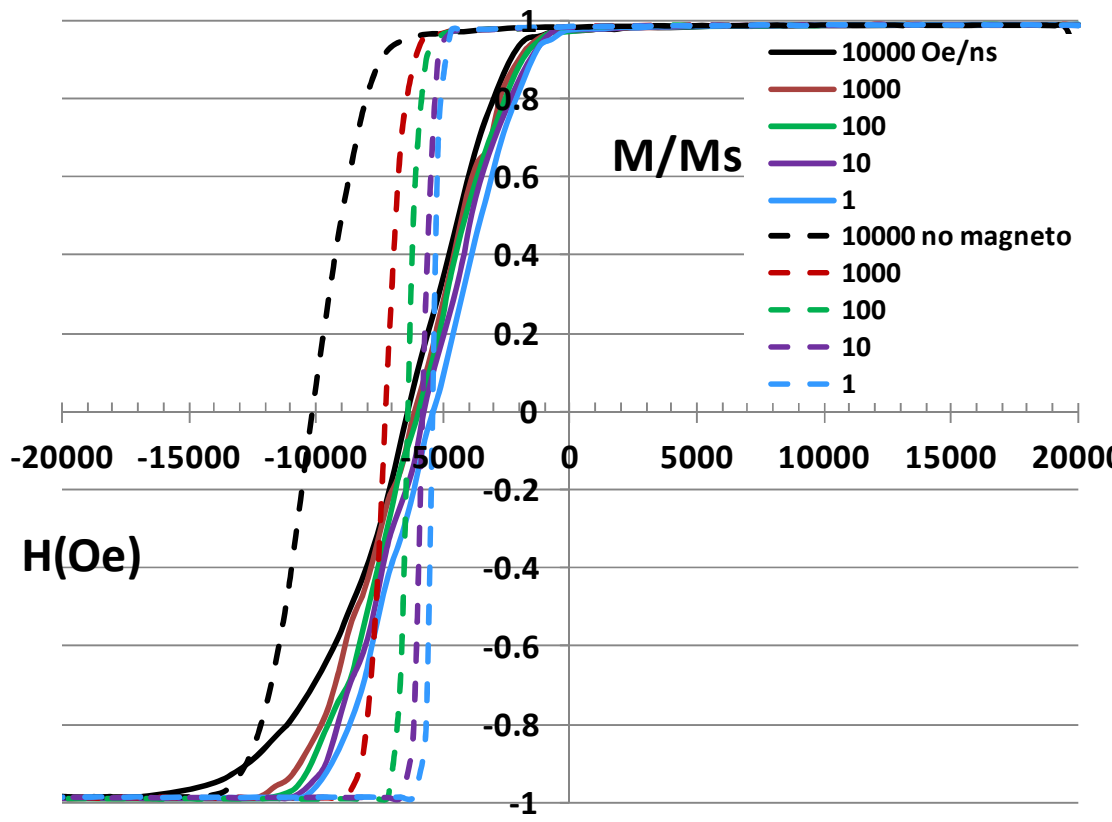

Fig. 7. ECC media with the field normal to the plane at 300 K with and without magnetostatic interactions included. Five different sweep rates were used, as indicated.

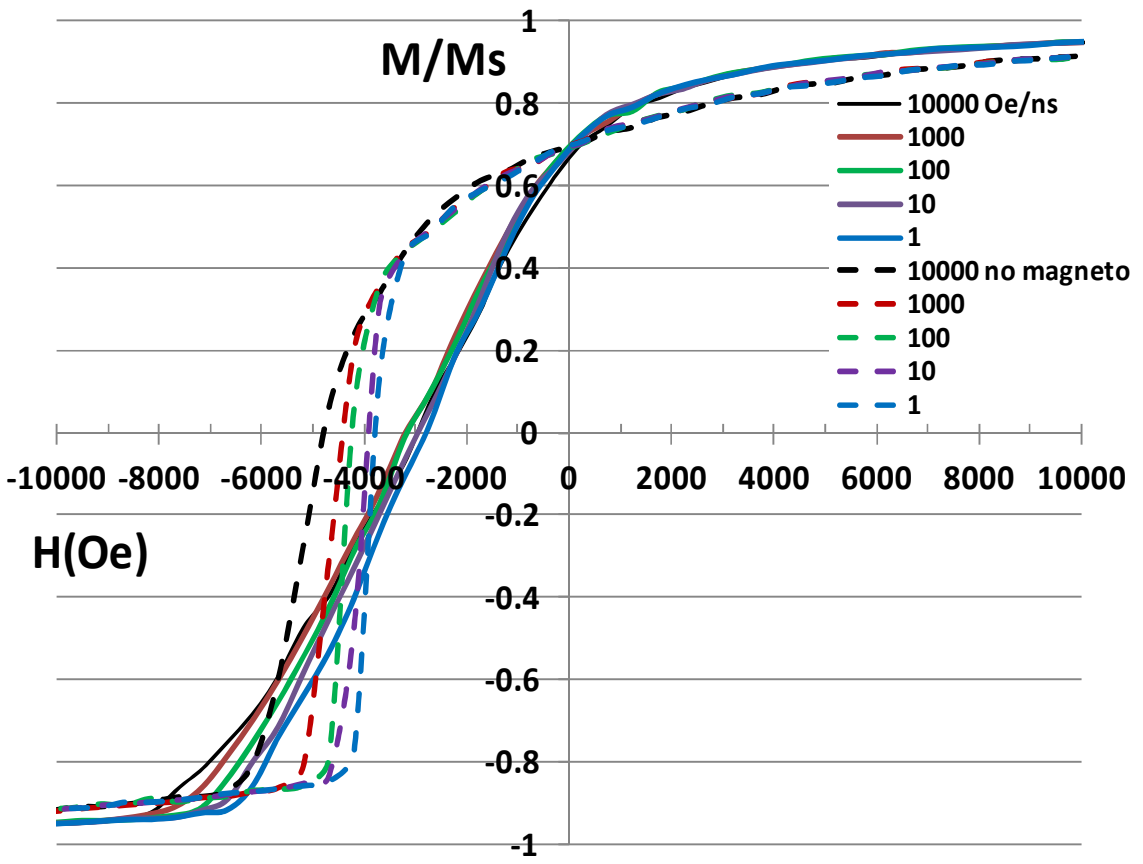

Fig. 8. ECC media as in Fig. 7 but with the applied field at $45^0$ to the media plane.

A summary of the sweep-rate dependence of $H_c$ corresponding to the ECC-media loops of Figs. 7 and 8 is given in Fig. 9. As in Figs. 3 and 4, the data exhibit near logarithmic behavior over the entire range of R values, except in the case of no magnetostatic interactions with the field along the easy axis. Another notable feature is the large difference in $H_c$ values with and without magnetostatic interactions only with the field at $45^0$. For single-layer media, a significant difference occurs at both directions of the applied field.

A comparison between the sweep-rate dependence of $H_c$ for single-layer and ECC media is shown in Fig. 10 at the four values of T and with all interactions included. A major conclusion from these results is that ECC media shows significantly less dependence on R than the single-layer case, consistent with expectations.[7,8,14] Although it is difficult to conclude with this limited data, the results on ECC media appear to show more non-linearity, at least at the lower temperatures, than the corresponding curves for the single-layer version. This point is quantified in the next section. Shallower slopes are also seen in the ECC curves of Fig. 11 with the field at $45^0$, although there is less of a difference with the single-layer case than in Fig. 10.

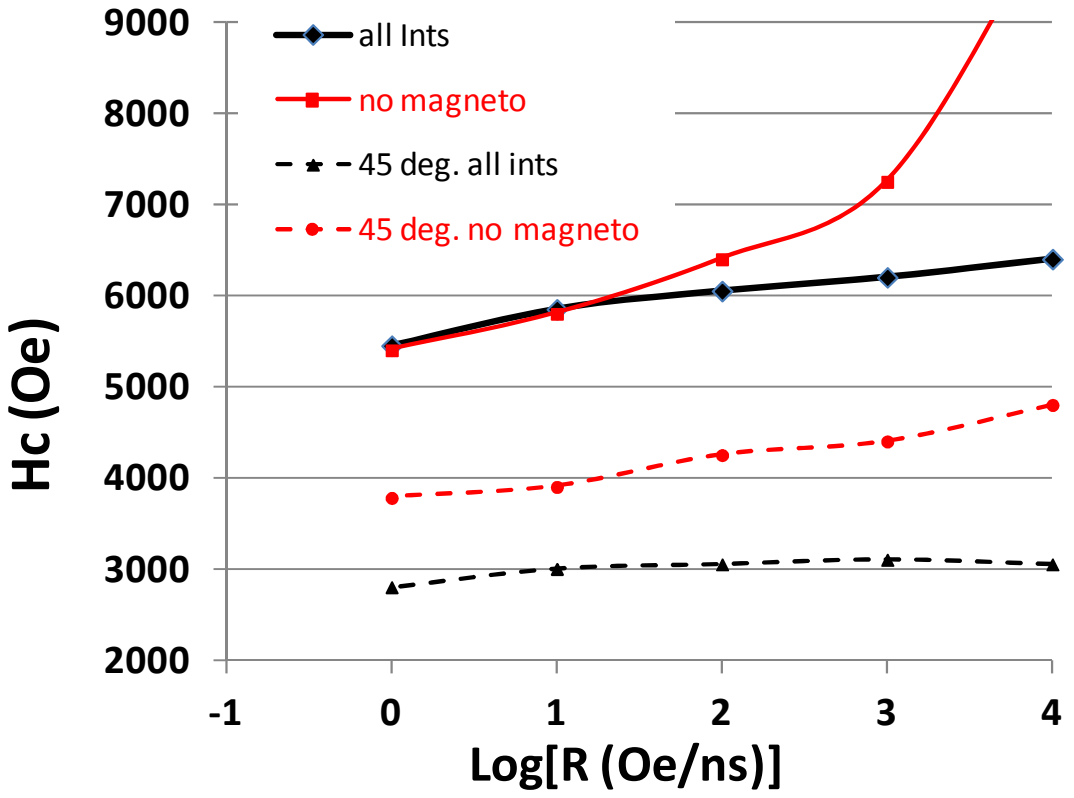

Fig. 9. Summary of the dependence of $H_c$ at T = 300 K on sweep rate for ECC media from the loops of Figs. 7 and 8.

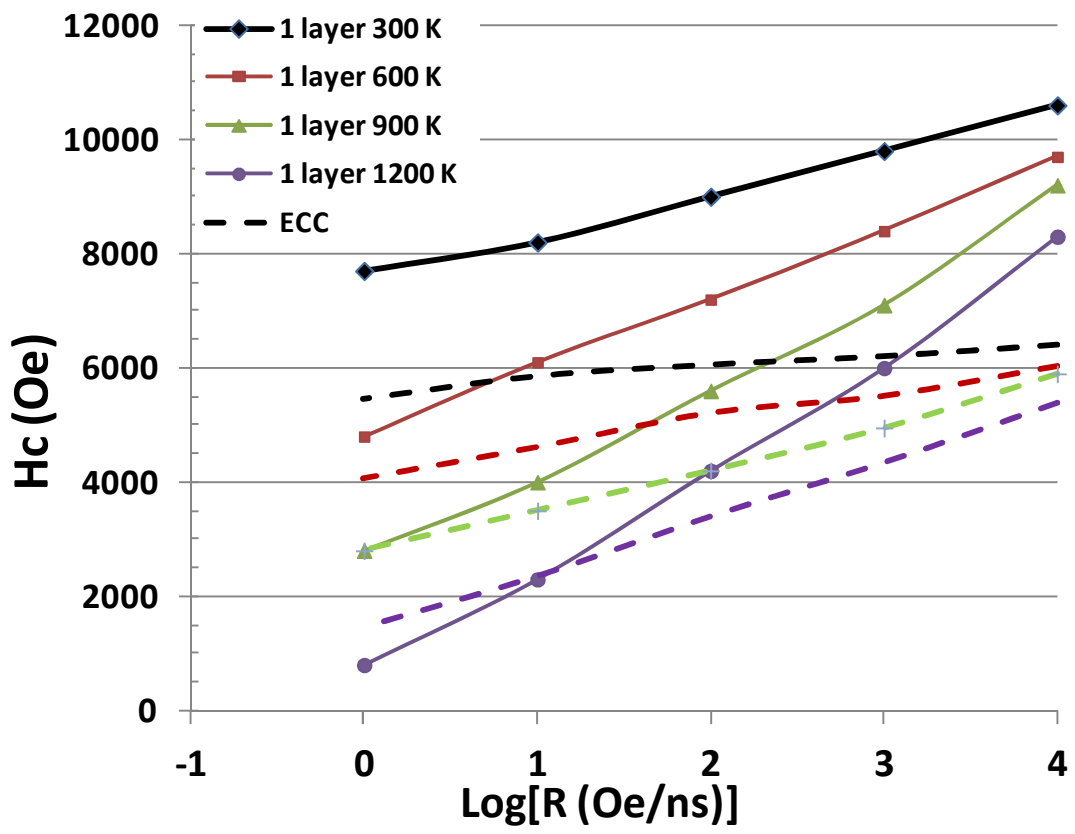

Fig. 10. Comparison of sweep-rate dependence of $H_c$ between single-layer (solid lines) and ECC media (broken lines) with the field normal to the plane. All interactions included.

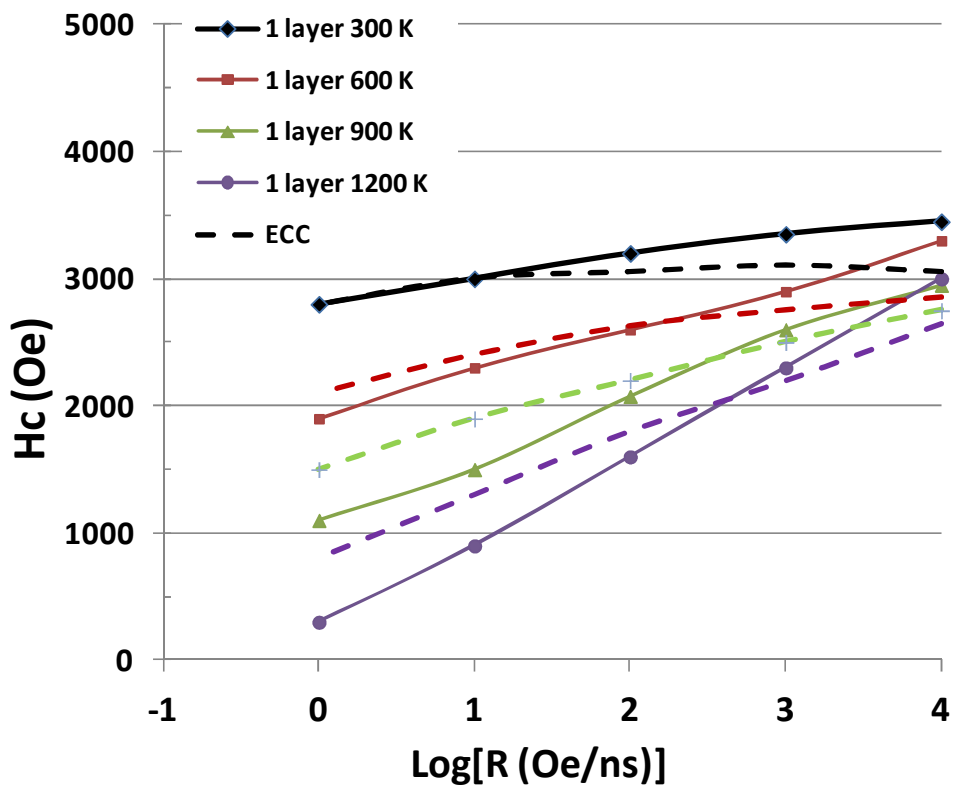

Fig. 11. As in Fig. 10 but with the field at $45^0$ to the media plane.

## V. FENG-VISSCHER SCALING

Using a switching probability rate equation, Feng and Visscher[5] derive a closed-form set of equations for $H_c(T,R)$. The analysis is based on the assumption of identical non-interacting grains subject to a uniaxial anisotropy K with thermally activated switching governed by the Arrhenius law

$$r = f_0 e^{-E/k_B T}. \tag{1}$$

involving the attempt frequency $f_0$ and energy barrier

$$E = KV(1 - H/H_0)^n, \tag{2}$$

where V is the grain volume, $H_0=H_K=2K/M_s$ in the case where the field is along the magnetization axis. The exponent n is usually taken to have the value 2 if the field is aligned along the easy axis but can vary from 2 to about 1.5 depending on the field direction.[16] Their analysis leads to the relation between $H_c$ and R and T given by the following set of equations:

$$R = R_0 erfg_n[s(1 - H_c/H_0], \tag{3}$$

where

$$R_0 = \frac{\sqrt{\pi} f_0 H_0}{2sln2} \tag{4}$$

$$s = (KV/k_B T)^n \tag{5}$$

$$erfg_n(y) = \frac{2}{\sqrt{\pi}} \int_y^{\infty} e^{-z^n} dz \tag{6}$$

$$y = s(1 - H_c/H_0). \tag{7}$$

These relations were shown to reduce to known results in various limits. The authors successfully fit these formulae to micromagnetic simulation results on a model recording media using a number of values for K but over a limited range of rather large R-values, from 2.5 – 40 kOe/ns, and with s and $R_0$ as fitting parameters, n is constrained to the value 3/2, and $H_0$ constrained to be $H_K/2$. The temperatures at which the simulations were performed were not stated.

We employ a somewhat different application of Eqs. (3) – (7) to fit to the micromagnetic results of the previous two sections. First, we use $H_K=2K_1/M_s$ (with $K_1$ given by the hard layer value in the ECC media case) and the Stoner-Wolhfarth (SW) relation between $H_0$ and $H_K$ for a given angle φ between the easy axis and field direction,[16]

$$H_0 = H_K[sin\varphi^{2/3} + cos\varphi^{2/3}]^{-3/2}. \tag{8}$$

In this work, φ is set to $0^0$ or $45^0$. However, we have also extended this result to account for the $3^0$ angular distribution in the grain easy-axis directions. This reduces $H_0$ by factors of 0.86 and 0.50 in the cases of $\varphi=0^0$ and $45^0$, respectively.[18] Next, we adopt the formulation of Wood[17] to set the value of the exponent n where an approximate formula is given (for a single SW grain) as a function of φ, which we have also extended to account for the angular.[18] When this is done, we find n=1.78 for $\varphi=0^0$ and n=1.42 for $\varphi=45^0$. A best fit of the data (for each configuration described in Sects. III and IV), for the five values of R at each of the four values of temperature, was then made by adjusting $f_0$ individually for each temperature (yielding four fitted values for $R_0$). For ECC media, only the hard-layer parameter values were used in these formulae. If the concepts of SW switching Arrhenius scaling are applicable, the data from all 20 values of $H_c$ in each case studied should collapse onto a single curve.

Fig. 12 shows the results of such a procedure in the single-layer case with all interactions included and the field normal to the plane ($\varphi=0^0$). The procedure yielded $f_0$ values 23.9, 17.7, 16.4, and 18.0 GHz (with increasing T), giving an average $f_0$=19 GHz. This value is consistent with analytic estimates[19] and the scaling theory appears to fit the data well. Reasonably good fits were also obtained in the other cases studied (with and without exchange and magnetostatics) for single-layer media with $\varphi=0^0$ but resulting in very different $f_0$ values for each case. For example, with no interactions, the average $f_0$ was found to be 6.4 GHz with $R^2$=0.999.

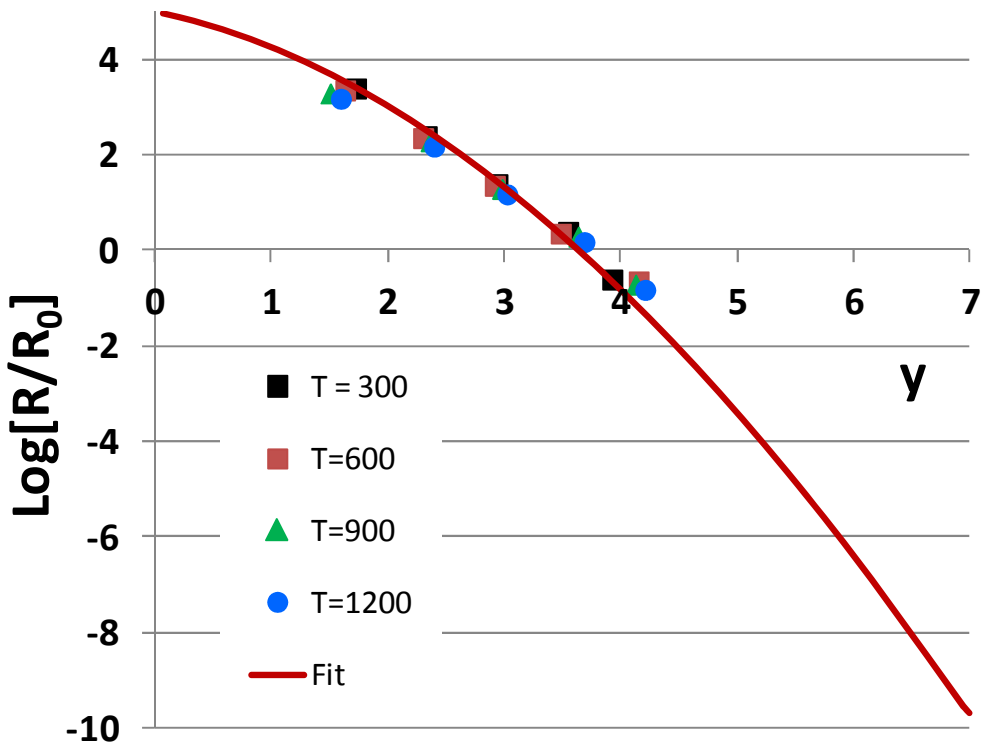

Fig. 12. Fit of simulated coercivity values for the field normal to the plane in single-layer media (all interactions included) using the scaling model of Feng and Visscher.[5]

Significantly less satisfactory fitting of the micromagnetic results to this Arrhenius-type scaling was found in the case of the field applied at $45^0$, as shown in Fig. 13. The resulting $f_0$ values spanned an exceedingly unphysical range from 2.7 $x10^{10}$ GHz to 550 GHz, suggesting that the simple assumption of the model are not applicable in this case.

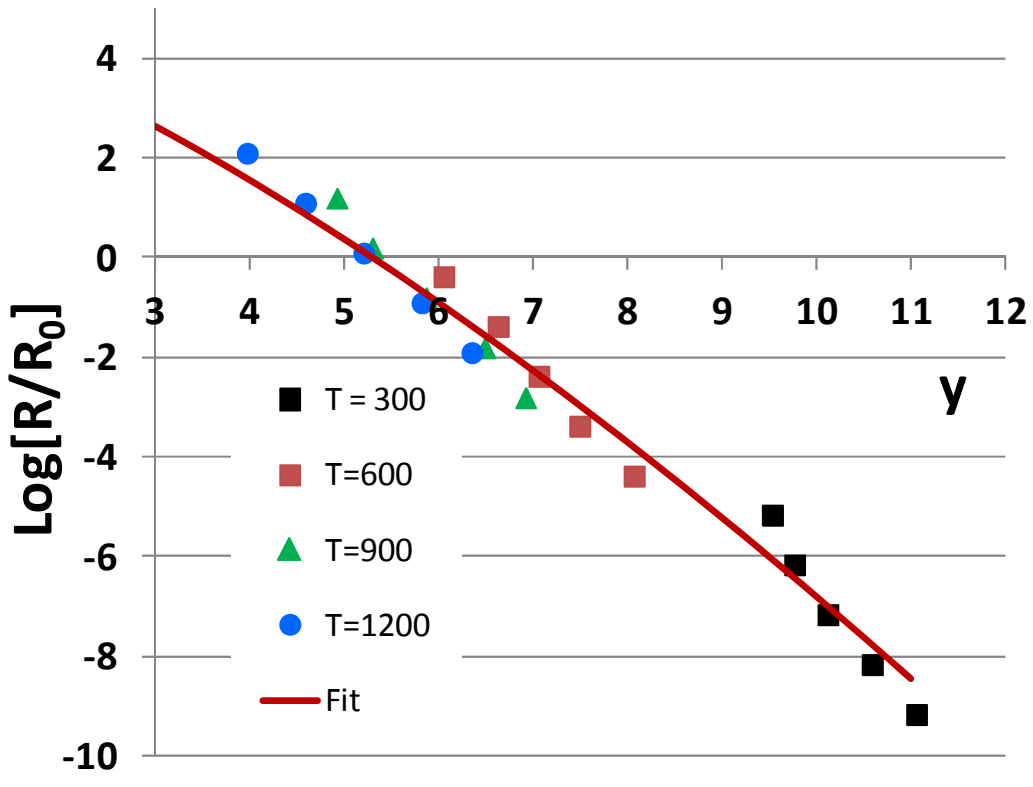

Fig. 13. As in Fig. 12 but with the applied field at $45^0$ to the single-layer media plane.

A similarly poor fit was determined in the case of ECC media. For the field applied normal to the plane, the resulting $f_0$ values again were unphysical, ranging from 6.3 x$10^6$ GHz to 58 GHz. Fig. 14 shows the resulting fit of all the data to the Feng-Visscher relation.

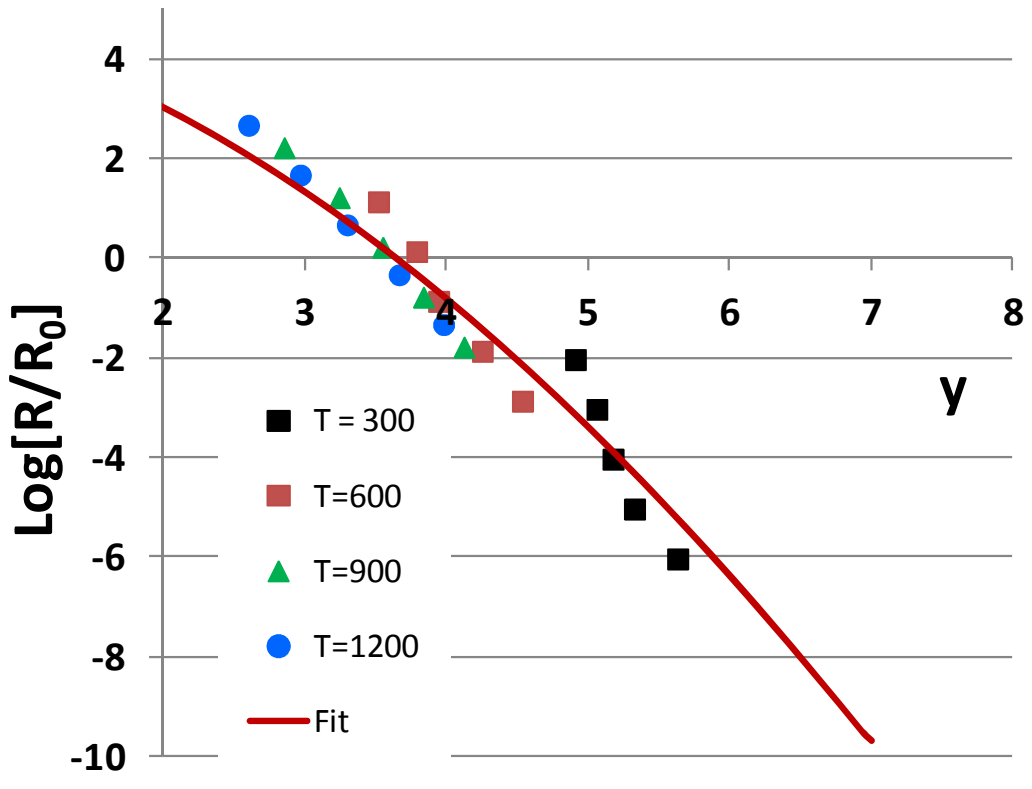

Fig. 14. As in Fig. 12 but for ECC media with the field applied normal to the plane.

Not surprisingly, the worst case was for ECC media with the field applied at $45^0$, as shown in Fig. 15, exhibiting little relationship with the simple scaling assumptions. Fitted $f_0$ ranged from 1.1 x$10^{11}$ GHz to 380 GHz.

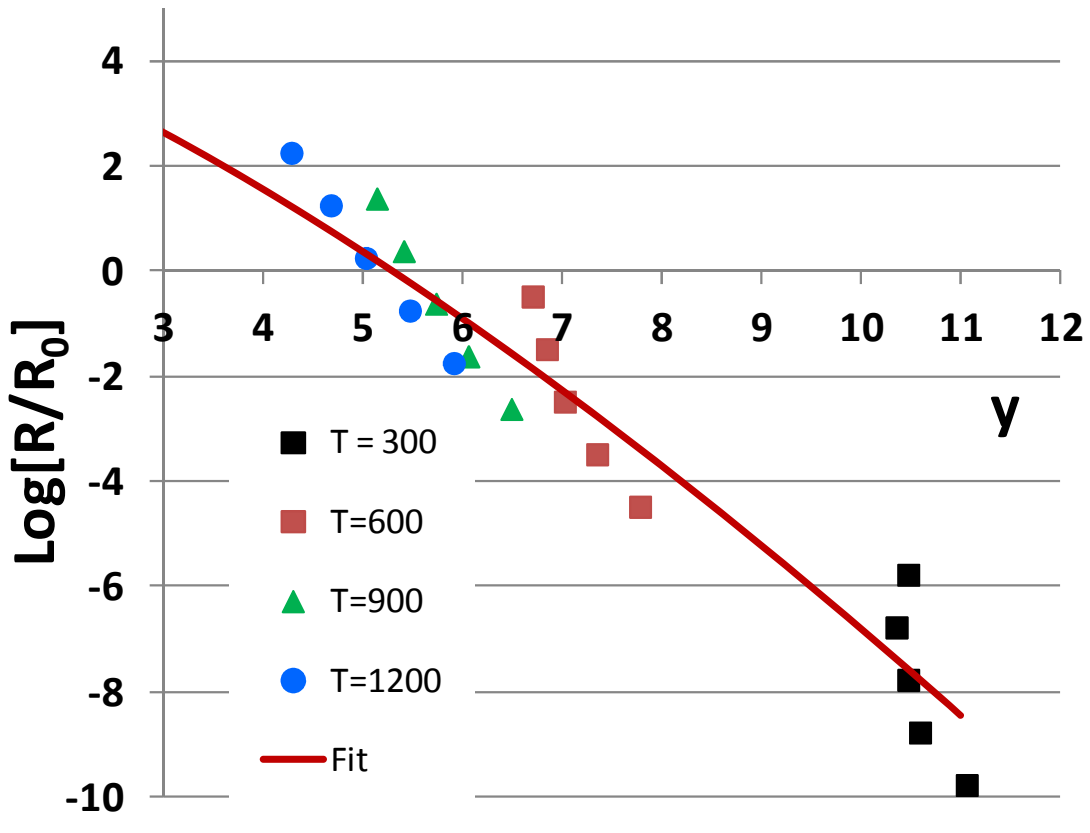

Fig. 15. As in Fig. 12 but for ECC media with the field at $45^0$.

## VI. Discussion and Conclusion

In this work we have explored thermally activated magnetic grain switching through the study of the temperature and sweep-rate (R) dependence of the coercivity for high anisotropy perpendicular recording media. Micromagentic simulations of M-H loops, which included distributions in the various material parameters, were performed on both single-layer and ECC with the field applied either normal to the plane or at $45^0$ at four temperatures and five sweep rates. One conclusion of these results is that the reduction of $H_c$ is roughly logarithmic in R for each case considered and is largely unaffected by inter-grain interactions. We also find significantly weaker dependence of $H_c$ on R with the field at $45^0$. A comparison of the two types of media shows that $H_c$ in dual-layer case exhibits notably less reduction as the sweep rate is made slower than for single-layer media. This reduced sensitivity implies an enhanced robustness of grain magnetization switching to thermal fluctuations.

We have also fit our simulation results for $H_c(T,R)$ to the Arrhenius-Néel/Sharrock type scaling model proposed by Feng and Visscher.[5] This procedure involved treating the attempt frequency $f_0$ as a fitting parameter at each of the four simulation temperatures. In the case of single-layer media with the field applied normal to the plane this yielded a reasonably good fit to the numerical data for all temperatures collapsed onto a single scaling curve, as predicted by the theory. For the field at $45^0$ and for ECC media, the fits gave unphysical values for $f_0$ and global scaling was unsatisfactory.

The characterization of the dynamics of thermally assisted magnetization reversal remains a significant challenge but also a worthwhile goal. A more reliable means by which to use long-time experimental room-temperature M-H media loops as means by which to fit short-time microscopic modeling parameters would be valuable for use in recording models.[18] Full characterization requires agreement between the modeled results and experimental loops for the field applied in any direction. Simulations of the longer-term decay of recorded media transitions is even more of an issue. The present work has demonstrated that the use of procedures based on Arrhenius-Néel-type scaling to extract long-time behavior from short-time simulations may be useful only for single-layer media described in terms of a simple anisotropy-dominated energy barrier.[9] This may not be surprising given underlying assumptions of the Arrhenius law. Alternative approaches to brute-force micromagnetic modeling and simple scaling assumptions that are based on extensions of the SW model with Monte-Carlo-type simulations of switching appear to be more promising.[6,18]

## Acknowledgments

This work was supported by the Natural Science and Engineering Council (NSERC), The Canada Foundation for Innovation (CFI), and Western Digital Corporation.

[1] *The Physics of Ultra-High-Density Magnetic Recording*, Eds. M. Plumer, J. van Ek and D. Weller (Springer-Verlag, Berlin, 2001).

[2] M.P. Sharrock, IEEE Trans. Magn., 20, 754 (1984).

[3] R.W. Chantrell, G.N. Coverdale, and K. O'Grady, J. Phys. D: Appl. Phys. 21, 1469 (1988).

[4] M. El-Hilo, A.M. de Witte, K. O'Grady, and R.W. Chantrell, J. Magn. Magn. Mater. **117**, L307 (1992).

[5] X. Feng and P. B. Visscher, J. Appl. Phys. **95**, 7043 (2004).

[6] P.-L. Lu and S.H. Charap, J. Appl. Phys. 75, 5768 (1994).

[7] D. Suess, S. Eder, J. Lee, R. Dittrich, J. Fidler, J.W. Harrell, T. Schrefl, G. Hrkac, M. Schabes, N. Supper, and A. Berger, Phys. Rev. B **75**, 174430 (2007).

[8] Saharan, C. Morrison, J.J. Miles, T. Thomson, T. Schrefl, G. Hrkac, J. Appl. Phys. **110**, 103906 (2011).

[9] J. Xue and R.H. Victora, Appl. Phys. Lett. **77**, 3432 (2000).

[10] http://llgmicro.home.mindspring.com/

[11] M.L. Plumer, M.C. Rogers, and E. Meloche, IEEE Trans Magn. **45**, 3942 (2009).

[12] M.L. Plumer, J. van Lierop, B.W. Southern, and J.P. Whitehead, J. Phys.: Condens. Matter **22**, 296007 (2010).

[13] J. I. Mercer, M. L. Plumer, J. P. Whitehead, and J. van Ek, Appl. Phys. Lett. **98**, 192508 (2011).

[14] J.-P. Wang, W. Shen, and S.-Y. Hong, IEEE Trans. Magn. **43**, 682 (2007).

[15] *Introduction to Magnetic Materials, 2nd ed.* , B.D. Cullity and C.D. Graham (Wiley, Hoboken, 2009).

[16] R.H. Victora, Phys. Rev. Lett. **63**, 457 (1989); J.W. Harrell, IEEE Trans. Magn. **37**, 533 (2001).

[17] R. Wood, IEEE Trans Magn. **45**, 100 (2009).

[18] T. J. Fal, M. D. Leblanc, J. P. Whitehead, M. L. Plumer, J. I. Mercer, and J. van Ek (unpublished).

[19] X. Wang and H.N. Bertram, J. Appl. Phys. **92**, 4560 (2002).